\documentstyle[prl,aps,multicol,epsf]{revtex}

\begin{document}
\draft
\title{Magnetism Controlled Vortex Matter}
\author{Igor F.Lyuksyutov$^{a,*}$ and Valery Pokrovsky $^{a,b}$}
\address{(a) Department of Physics, Texas A\&M University,\\
College Station, TX 77843-4242 \\
(b) Landau Institute for Theoretical Physics, Moscow, Russia}
\date{\today}
\maketitle

\begin{abstract}
\noindent We discuss 
a new class of phenomena based on strong interaction
between magnetic superstructures
and vortices in superconductors in combined heterogeneous
structures.
An inhomogeneous magnetization 
can pin  vortices or 
create them spontaneously changing drastically 
properties of the superconductor.
On the other hand, the interaction between magnetic
moments mediated by vortices can result in specific
types of magnetic ordering. The same interaction
can create coupled magnetic-superconducting defects.
We discuss possible  experimental observation
of magnetism controlled vortex matter
in  superconducting films with magnetic nanoscale
dots or stripes and  layered  systems
with alternating superconducting and 
magnetic layers. 

\end{abstract}

\begin{multicols}{2}
\narrowtext
The purpose of this article is to indicate a new class
of heterogeneous systems realizing strong interaction 
between superconductivity(SC) and ferromagnetism(FM)
and to consider some physical effects in these systems. 
The interaction of superconductivity and ferromagnetism
has long been a subject of theoretical and experimental study
(see e.g. reviews \cite{bulaevskii,fisher}).  
In conventional systems  
both order parameters were 
homogeneous in space. In this case 
they suppress each other,
at least for singlet pairing.
As a result, one or both ordering is weak.
Recently we   proposed to employ heterogeneous 
structures which allow to 
enhance  drastically the SC-FM interaction 
\cite{dotlp},\cite{spie}.
By separating the magnetic and superconducting systems 
in space it is possible to avoid 
their mutual suppression due to the proximity effect. At the same time
the magnetic field induced by the nonuniform magnetization 
penetrates into the superconductor through 
superconducting vortices (SV) 
providing strong coupling between the  
vortex matter 
and magnetic textures.

A magnetic texture and  
superconducting vortex array (SVA) interact 
most effectively if
their characteristic scales are of the
same order of magnitude. The smallest scale 
which characterizes the SVA is the coherence length
$\xi$ which varies in a range 
Another characteristic distance is the London
penetration depth $\lambda$ which
can be in the range of $60-1000 nm $.
If the magnetic system has textures with characteristic
length $\xi < L < \lambda $, one can expect strong
interaction between the SVA and magnetic subsystem.
Such textures can be formed by topological defects
in the magnetic film (domain walls (DW), 
magnetic vortices (MV), skyrmions),
by artificial superstructures like arrays of 
magnetic dots on a superconducting film or by layered
magnetic/superconducting systems.
It is important to separate FM and SC layers
by insulating barriers to eliminate
the proximity effect and spin diffusion.

Below we discuss several possible heterogeneous SC-FM
systems. Consider an array of magnetic 
nanoparticles on a superconducting
film. The parameters of such a system can 
be varied in the process of growth and by the strength and
direction of  external magnetic field.
Each magnetic dot can create a single vortex (or a bunch of 
vortices) which can either 
increase vortex pinning, or
induce a resistive state, 
depending on the order in
the magnetic subsystem
and on the applied magnetic field.
The experimental investigations of magnetic dots
interacting  with 
superconducting films have been started.
So far  only the effect of pinning on the 
flux flow resistance and the critical
currents  was studied
\cite{spmsi1,shull,spmsi3,giv,barb}.  
In the previous work \cite{dotlp}  we predicted unusual
phase diagrams, spontaneous formation of the vortex plasma  
of unbound vortices in the case of dots
with magnetization normal to the film.
These phenomena occur if the coercive force of a
magnetic material used for dots is strong enough.
If dots are made from a soft magnet, the stable
state will be totally different: magnetic moments of 
the dots will be organized in a stripe domains with
opposite magnetization due to 2D dipolar interaction
between them mediated by vortices.

Another way to produce 
strong coupling between magnetism and vortices 
is to create  
{\it topological} magnetic defects like magnetic vortices, 
domain walls and skyrmions \cite{skyrm}, \cite{polyakov},
\cite{AP} in magnetic films.
Consider a thin superconducting film 
with a magnetic overlayer on the top of it.
The magnetic field outside the magnetic layer with
magnetization parallel or perpendicular 
to the film drops rapidly
at a distance of few {\AA} from the  film. However,
inhomogeneous magnetic field from a topological
magnetic defect propagates at a distance comparable with
the topological defect size resulting in  strong
coupling between magnetic and superconducting defects.
For domain walls this
size can be many tens of nanometers, 
larger than a typical superconducting vortex core
size. The skyrmion size is not fixed 
in the isotropic magnet \cite{AP}
and can be modified by the superconducting vortex. 
Experimentally, the creation of a skyrmion or a magnetic  vortex 
in the magnetic film is a challenging problem,
which can be alleviated for coupled skyrmion-vortex 
excitation. Details will be published 
elsewhere.

Layered systems  of superconducting 
films separated by textured layers of ferromagnetic 
materials can also provide strong interaction
between magnetization and vortices. The texture can
consist of dots or stripes to increase
the penetration of magnetic field into superconductor.
Some layered superconducting systems 
have been studied experimentally, 
and magnetic layered systems with
non-superconducting metallic layers up to 10 nm thick 
between the flat  magnetic layers have been used recently as
``spin valves'' in magneto-electronic devices.
Below we discuss some manifestations of
strong coupling between vortex matter and magnetism.

\noindent{\bf Magnetic/Superconducting Bi-layers}.
Consider first 
coupled SV-MV and SV-DW systems for a superconducting
film covered by a homogeneous magnetic overlayer.
   
\noindent{\it Superconducting/Magnetic Vortex Pair}.
The Hamiltonian of the  magnetic layer reads:

\begin{equation}
{\cal H}_M = \frac{J_M}{2}\int{d{\bf r}}({\nabla \bf m})^2
- \mu_B g_M\int d{\bf r}{\bf m}({\bf r}){\bf h}({\bf r}) 
\label{h1} 
\end{equation}
where ${\bf m}({\bf r})$ is the unit 2D vector
directed along the local magnetization,  
and ${\bf h}({\bf r})$ is the scaled magnetic field. 
The factor $g_M = C_M n a^{-2}$ where  
$n$ is the number of atomic layers in the magnetic film,  
$a$ is a lattice constant and $C_M$ is a numerical factor.
A single MV is characterized by the magnetization 
${\bf m} = {\bf r}/r $. 
There is a freedom of the global rotation of
all spins by the same angle.
The  MV energy  grows logarithmically
with the system size $L$ $E_{MV} \approx \pi J_{M}\ln({L}/{\xi})$.
However, the vortex-antivortex pair has a finite energy.
Dissociation of these pairs leads to the 
Berezinskii-Kosterlitz-Thouless (BKT) transition 
\cite{berez,KT} with  the transition temperature 
$T_{MV} = \pi J_{M}/2$ (see reviews \cite{nel,min}).
The energy of a single SV without magnetic field is 
(see e.g. \cite{Abr}):
\begin{equation}
 E_{SV} = 
\epsilon_0 \ln{l \over \xi} 
 +  \epsilon_{core}
\label{sv} \end{equation}
where $\epsilon_0 = {\phi^2_0 / 16 \pi^2  \lambda_{eff}}$,
$\epsilon_{core} \sim \epsilon_0$
is the energy of the  SV core
 and  $l$ is the characteristic cut-off length.
For a single SV the length $l$ is equal to 
the effective screening length 
$\lambda_{eff} = \lambda^2/d$, where
$\lambda$ is the London penetration 
depth and $d$ is the film thickness
(\cite{Abr,Abr64}).
In a specially prepared film $\lambda_{eff}$
can be comparable with the system size \cite{hebard,donhu}.
In this case the BKT transition 
\cite{berez,KT} with  the transition temperature 
$T_{SV} = \epsilon_0/2$  occurs (see reviews \cite{nel,min}).
If $\lambda_{eff}$ is not too very large the SV energy
is finite, and the SV can be excited 
thermally. The thermally excited SV lead to the 
resistive state of SC film at $T > 0$. 

The magnetic field from a SV has normal and parallel to
the film components. The total flux from the normal
component is equal to the  magnetic flux quantum
 $\phi_0$. 
The in-plane component is directed radially. 
Interacting with the 
magnetic vortex (MV) in the magnetic film,
it distinguishes the radial alignment of
the magnetization from other possible 
alignments. The in-plane magnetic field from a single SV
outside a thin superconducting film can be represented as 
${\bf h}=-\nabla \Phi  sign z$,
where function $\Phi (r)$ has a following form \cite{Abr}:
\begin{equation}
 \Phi (r,z) = { \phi_0  \over {2\pi}} 
\int^{\infty}_0 J_0(qr) \exp({-q \vert z\vert})
\frac{dq}{1+2q\lambda_{eff}}
\label{phi} \end{equation}
Here $ J_0(x)$ is the Bessel function.
At  small distances $r\ll \lambda_{eff}$ 
the magnetic field at a film is
$ {\bf h}(r,z=0) \approx {\phi_0{\bf r}/(2\pi r^2\lambda_{eff})}$.
At large distances $r\gg \lambda_{eff}$
the  in-plane magnetic field decreases as  
$ {\bf h}(r,z=0) \approx {\phi_0}{\bf r}/({2\pi r^3})$ \cite{Abr}. 
Note, that the normal to the film component $h_z$ decays as
$1/r^3$ for  $r\gg \lambda_{eff}$, 
faster than the in-plane component. 

Neglecting the thickness of layers and
the change in the SC current distribution
caused by magnetic field of the MV, 
we can calculate the interaction energy
of the two vortices assuming that their
centers are located exact by opposite each other.
It is possible to distinguish
two contributions to the 
MV energy due to the SV magnetic field.
The first one originates from the
region $r < \lambda_{eff}$.
It does not depend on $L$ and is equal to 
$ -g_M \mu_B \phi_0a^{-2} = -\pi J_{\phi}$. 
The second one  originates from the
region $r > \lambda_{eff}$.
It depends  logarithmically on $L$ and is equal to
$-\pi J_{\phi}\ln({L}/\lambda_{eff})$. 
The product $ \mu_B \phi_0 /\pi  = \hbar^2/(2m_e) = 3.8 eV {\AA}^2$.
The interaction between MV and SV is purely magnetic.
Its density is proportional to  $ 1/c^2$. But it is enhanced
due to its  long-range: $\lambda_{eff}\propto c^2$.
As a result the effective coupling constant   
$J_{\phi}/n$ has a typical atomic scale.
This energy is much larger than the energy of
the dipole-dipole interaction which, therefore, can be neglected.
The above calculation neglects the influence of the field
produced by the MV onto the supercurrent distribution. 
However, this calculation gives a reasonable estimate
of the interaction energy.
The interaction with SV magnetic field can make
MV energetically favorable.
For the energy of the coupled  MV-SV system we find:
\begin{equation}
E_{SM} = \pi(J_M - J_{\phi})\ln(\frac{L}{\xi})+ 
E_{SM0}
\label{sm} 
\end{equation}
where 
\begin{equation}
E_{SM0}=
\pi J_{\phi}\ln\frac{\lambda_{eff}}{e\xi}
+\pi J_{M}\ln\frac{\xi}{a}
+\epsilon_0 \ln{\frac{\lambda_{eff}}{ \xi}} 
 +  \epsilon_{core}
\label{sm0} 
\end{equation}
The most interesting is the case  $J_M < J_{\phi}$
easily achievable in the experiment. In this case the 
energy of the coupled pair MV/SV is negative and grows
with the system size. It leads to the first order transition
to a new phase with proliferated coupled MV/SV.
However, it is not yet clear whether high energy 
barrier for this process can be overcome for reasonable 
time. Magnetic field perpendicular to the film can alleviate
the activation process.    
When $J_M > J_{\phi}$, the superconducting/magnetic vortex (SMV)
energy diverge logarithmically with the system size.
The  energy of the  vortex-antivortex pair 
is finite. Thus, the system undergoes BKT 
transition \cite{berez,KT,nel,min}
with  the transition temperature 
$T_{2D} = (J_M-J_{\phi}) /2 $. If the bulk superconducting
temperature $T_S$ is in the range $T_{2D} < T_{S} < T_{MV}$,
the BKT transition  at $ T = T_{MV}$
into the  state with the algebraic magnetic order.
Below  $T_{S}$ bound superconducting/magnetic 
vortices proliferate. They
destroy the magnetic order in the magnetic film
in the temperature interval  $T_{2D} < T < T_{S}$.
At temperature below  $T_{2D}$ the SMV 
are thermodynamically unfavorable and 
the algebraic magnetic
order revives in the magnetic film.

\noindent{\it Domain Wall and Superconducting Vortices}. 
Consider a ferromagnetic film with in-plane magnetization
and  easy-plane in-plane uniaxial anisotropy.
Its Hamiltonian has a form:
\begin{equation}
{\cal H}_M = \int{d{\bf r}}\left( \frac{1}{2}J_M({\nabla \bf m})^2
-\lambda_i ({\bf m}({\bf r})\cdot{\bf n}_{\parallel})^2\right) 
\label{h2} 
\end{equation}
where ${\bf m}({\bf r}) $ is a 2D unit vector field, $\lambda_i$ 
is the in-plane anisotropy constant and ${\bf n}_{\parallel}$ is the
unit vector along anisotropy axis.
This system has the Ising symmetry. The domain wall width is 
$l_{DW} \propto a \sqrt{J_{M}/\lambda_i}$ 
and the  domain wall energy per unit length is 
$E_{DW} \propto a \sqrt{J_{M}\lambda_i}$. 
The inhomogeneous magnetization in the domain wall
generates magnetic field which penetrates
into the superconducting film in a stripe of the width $l_{DW}$. 
The energy of the SV in the magnetic field created by 
the domain wall decreases:
\begin{equation}
E_{SV}=
\epsilon_0 \ln{\frac{\lambda_{eff}}{ \xi}} 
-\frac{\pi}{2}M{\phi}_0\frac{l^2_{DW}}{\lambda_{eff}}
\label{dw} 
\end{equation}
where $M$ is the magnetization of the ferromagnet. 
$E_{SV}$ is negative if 
$M>M_0 = {\phi}_0\ln{{\lambda_{eff}}/{ \xi}} /(8\pi^3 l^2_{DW})$.
Thus, if $M>M_0 $ the vortices  proliferate until
their interaction in chain stops this process.
The equilibrium density of vortices in the 
chain is equal to 
$(\ln{{\lambda_{eff}}/{ \xi}}/{\lambda_{eff}})(M/M_0-1).$
The linear tension of the domain wall in the presence
of the SV is equal to 
\begin{equation}
E^{\prime}_{DW}=\pi\sqrt{J_{M}/\lambda_i}-E_{SV}n 
= E^{(0)}_{DW}n-\vert E_{SV}\vert n 
\label{dwsv} 
\end{equation}
It becomes negative if 
\begin{equation}
M/M_0 -1 > 
(E^{(0)}_{DW}{\lambda_{eff}}/\epsilon_0)^{1/2}
\ln^{-1}{{\lambda_{eff}}/{ \xi}}.
\label{dwsv1} 
\end{equation}
For estimates we put $\lambda_{eff} = 1\mu m$,
$ \xi =100$\AA, $\epsilon_0 = 0.4 eV$,
$l_{DW} = 100$\AA, $E^{(0)}_{DW} = 0.013 eV/$\AA,
$4\pi M_0 = 50 Oe$
Then the threshold value $4\pi M_{th}$ 
is about $400 Oe$. At $M > M_{th}$
the SC transition is accompanied by a simultaneous
magnetic transition to the disordered or 
heterogeneous phase with multiple domains.
Domain walls with the SV chains on them can
control the transport properties. They are easily 
regulated by the in-plane magnetic field. 

\noindent{\it Stripe Domain Structure} 
In the case when the magnetization is normal  to the film,
the stripe domain structure caused by dipolar fforces
was predicted \cite{dipole} and observed \cite{stripe}.
Magnetic field from this superstructure penetrates
into the superconducting film and can create a vortex array.
The presence of vortices can
drastically change the stripe structure.

\noindent{\it Current Driven Magnetization and Field Controlled Transport}. 
Electric current  larger than the critical one excites a
maltitude of  vortices and  
magnetic topological defects bound with them. 
Corresponding changes in magnetization 
can be observed experimentally. 
It is possible to observe reverse effect: a voltage generated
in superconducting film by applying
magnetic field  parallel to the
film.

\noindent{\bf Magnetic/Superconducting Multi-layers:}
Consider artificial layered
structures consisting of superconducting (SC)
layers alternating with corrugated or
textured  ferromagnetic(FM) layers.
The properties of  such  tri- and multi-layers
depend on layer thicknesses, 
magnetic layer textures and corrugations.
According to the concept of our article,
strong interaction between FM and SC layers is
provided by the magnetic field induced by the FM layers and
penetrating into SC layers. The magnetic field does not
propagate  outside ideally smooth magnetic layers with parallel
interfaces for both parallel and normal magnetization.
To overcome this difficulty, it is possible to 
corrugate the interface with a characteristic scale
of the corrugation  larger than SC coherence length.
It was shown experimentally \cite{step} that
even atomic size steps can orient the  magnetization. 
It is also possible to employ   films
with the modulated thickness 
or with a texture  of  magnetic islands 
or stripes deposited on the interface 
between the magnetic and superconducting layers.

\noindent{\it Ferromagnetic/Superconducting Sandwiches}
In  FM/SC/FM  sandwiches  with sufficiently thick 
superconducting layers the  exchange interaction
between magnetic layers is negligible.
In FM/SC/FM tri-layers made 
with soft magnet the induced SC vortices provide an
antiferromagnetic (AF) interaction
between magnetic layers. In similar  tri-layers made with
hard magnets the magnetization of ferromagnetic layers can
be  either parallel/metastable or antiparallel.
The magnitude of the in-plane critical current in the direction
perpendicular to the texture must be invariant on
reversal  of the current 
for parallel magnetization and must change
drastically for anti-parallel magnetization
(the sandwich is assumed to be symmetric with respect to the
central plane).   

\noindent{\it Magnetic superstructures 
on superconducting films:}
In the case of a regular structure of magnetic dots
with one or several flux quanta per dot, the flux line lattice
is strongly pinned.
When the magnetic flux 
associated with the dot is too small
to create vortices, it can nevertheless effectively pin
vortices. 

The magnetic moment of a magnetic film 
can be directed either parallel or 
normal to the surface of the superconducting film
in a controllable way \cite{all}. 
So far only particle arrays with magnetic moments parallel to the
film were studied experimentally 
 \cite{spmsi1,shull,spmsi3,giv,barb}.
The magnetic flux
from nanomagnets with magnetic moments 
aligned in the direction normal to the 
superconducting film can vary in a broad range, 
from a few percent up to many flux quanta.
The phase diagram of the  magnetic dot array
versus magnetic field and temperature depends on
coercive force of the dots, magnetic moment direction, 
geometry of the system (periods of the array, film thickness)
and superconducting properties of the 
film (London penetration depth).

The dots with magnetic moments 
parallel to the SC film
can bind vortex-antivortex  pairs.
Electric current in the SC film can orient
such a pair and in effect  rotate
the magnetic moment of the particle. 
To realize this effect  experimentally 
the pinning barriers should  be maximally reduced.
It can be done by choosing of soft
magnets for magnetic films, by giving them circular shape
and by controlling their homogeneity.

To conclude, we have shown
that by varying the system geometry, 
materials and magnetic field it is possible  to 
control the film resistivity, vortex motion,
and the dynamics  of nanoparticle magnetization.
We propose experiments to study 
the bound states of textured magnetic layers and vortices, 
and formation of coupled topological 
defects in magnetic and superconducting films, e.g. pairs of
skyrmion-SV, MV-SV, and domain walls-SV.

This work was partly supported by the grants
DE-FG03-96ER45598,
NSF DMR-97-05182, THECB ARP 010366-003. 
It is a pleasure to acknowledge discussions with 
D.Naugle, I.K.Schuller, J.Erskine, 
A. de Lozanne and B.Barbara.

\end{multicols}

\end{document}